\documentclass[prl,twocolumn,showpacs]{revtex4}
\usepackage{amssymb}
\usepackage{graphicx}
\usepackage{color}

\begin{document}

\newcommand{\be}{\begin{equation}}
\newcommand{\ee}{\end{equation}}
\newcommand{\nus}{\nu_{\mathrm{s}}}
\newcommand{\Omegas}{\Omega_{\mathrm{s}}}
\newcommand{\OmegaK}{\Omega_{\mathrm{K}}}
\newcommand{\tgw}{\tau_{\mathrm{GW}}}
\newcommand{\tdamp}{\tau_{\mathrm{damp}}}
\newcommand{\tsv}{\tau_{\mathrm{SV}}}
\newcommand{\tek}{\tau_{\mathrm{BL}}}
\newcommand{\Nacc}{N_{\mathrm{acc}}}
\newcommand{\Ngw}{N_{\mathrm{GW}}}
\newcommand{\Lacc}{L_{\mathrm{acc}}}
\newcommand{\Lgw}{L_{\mathrm{GW}}}
\newcommand{\Lheat}{L_{\mathrm{heat}}}
\newcommand{\Lnu}{L_\nu}
\newcommand{\Lmu}{\Lnu^{\mathrm{MU}}}
\newcommand{\Lsf}{\Lnu^{\mathrm{SF}}}
\newcommand{\Tc}{T}
\newcommand{\Tcp}{T_{\mathrm{cp}}}
\newcommand{\Tcn}{T_{\mathrm{cn}}}
\newcommand{\Tcnmax}{T_{\mathrm{cn,max}}}
\newcommand{\Lsflo}{\Lnu^{\mathrm{npSF}}}
\newcommand{\Lsfhi}{\Lnu^{\mathrm{pSF}}}
\newcommand{\ApJ}{Astrophys. J.}
\newcommand{\MNRAS}{Mon. Not. R. Astron. Soc.}

%%%%%%%%%%%%%%%%%%%%%%%%%%%%%%%%%%%%%%%%%%%%%%%%%%%%%%%%%

\title{Revealing the physics of r-modes in low-mass X-ray binaries}
\author{Wynn C.~G. Ho}
\email{email: wynnho@slac.stanford.edu}
\author{Nils Andersson}
\email{email: na@maths.soton.ac.uk}
\affiliation{School of Mathematics, University of Southampton,
Southampton, SO17 1BJ, United Kingdom}
\author{Brynmor Haskell}
\email{email: b.d.l.haskell@uva.nl}
\affiliation{Astronomical Institute `Anton Pannekoek,' University of Amsterdam,
1098XH Amsterdam, Netherlands}
\date{\today}

%%%%%%%%%%%%%%%%%%%%%%%%%%%%%%%%%%%%%%%%%%%%%%%%%%%%%%%%%
\begin{abstract}
We consider the astrophysical constraints on the gravitational-wave driven
r-mode instability in accreting neutron stars in low-mass X-ray binaries.
We use recent results on superfluid and superconducting properties
to infer the core temperature in these neutron stars
and show the diversity of the observed population.
Simple theoretical models indicate that many of these systems reside inside
the r-mode instability region.
However, this is in clear disagreement with expectations, especially for
the systems containing the most rapidly rotating neutron stars.
The inconsistency highlights the need to re-evaluate our understanding of
the many areas of physics relevant to the r-mode instability.
We summarize the current status of our understanding, and we discuss
directions for future research which could resolve this dilemma.
\end{abstract}


\maketitle

What limits the spin rate of a neutron star?
Given that the fastest rotating neutron star (NS) in a low-mass X-ray
binary (LMXB; binary system in which X-rays are produced when matter
is accreted onto the NS from a low-mass stellar companion)
and the fastest radio pulsar have spin rates
which are significantly below the centrifugal break-up limit,
it is natural to ask whether a physical mechanism
prevents further spin-up during the evolution of these systems.
The issue is challenging because of complex, often poorly understood,
physics.
One possibility is that the emission of gravitational radiation from the
NS plays a significant role.
Alternatively, the answer could be due to the detailed nature of the
accretion of matter (and angular momentum) onto the NS surface.
Despite its obvious importance, the question
remains unresolved, with only the current gravitational wave (GW) searches and
X-ray observations serving as constraints \cite{chakrabartyetal03}.

This Letter concerns one of the main mechanisms that is expected to affect
the spin evolution of an accreting star:
the instability associated with the r-modes, which are a class of
oscillations in a star whose restoring force is the Coriolis force.
The emission of gravitational waves can excite r-modes in the NS core
and cause the amplitude of the oscillations to grow.
The notion that this instability can provide a spin-limit for NSs in LMXBs
was first discussed in \cite{bildsten98}.
The r-mode instability is interesting for many reasons, mainly
because the associated gravitational wave signal may be detectable with
ground-based instruments, but also because its understanding
requires knowledge from a wide range of physics.
The primary agents that enter the r-mode discussion are
(1)  damping mechanisms related to the standard shear and bulk viscosities
and exotica like hyperons, quarks, and superfluid vortices, and
(2) the fluid dynamics associated with the mode, e.g., nonlinear
coupling and saturation \cite{anderssonkokkotas01,arrasetal03,bondarescuetal07}.
The instability depends primarily on the NS spin rate
$\nus$ and core temperature $\Tc$.
This leads to an instability ``window,'' determined by a critical curve
(defined by the balance of evolution timescales $\tgw=\tdamp$)
in the $\nus$-$\Tc$ plane, inside which the instability is active.
So far, most studies of the unstable r-modes focused on particular damping
mechanisms, in order to determine the extent to which they can kill
the instability or are relatively unimportant.
In several cases, e.g., hyperon bulk viscosity, the answer changed
as our understanding improved \cite{lindblometal99,haskellandersson10}.
For accreting NSs in LMXBs, the general view is that the
main damping mechanism is related to a viscous boundary layer at the
crust-core interface \cite{bildstenushomirsky00,levinushomirsky01}.
What has not been appreciated is that this model leaves the {\it majority} of
the observed LMXBs significantly inside the instability window:
rapidly rotating NSs should not possess spin rates at their observed levels.
This is the primary message of our Letter and one that requires attention
and resolution.

Previous discussions considered the general region in $\nus$-$\Tc$ where
the LMXB population resides \cite{anderssonkokkotas01,bildstenushomirsky00}.
Here we provide detailed estimates of the likely core temperatures for
each LMXB, accounting for nucleon superfluidity and superconductivity at
the level indicated by recent results from the NS in the Cassiopeia~A
supernova remnant (the youngest NS in the Galaxy)
\cite{pageetal11,shterninetal11}.
This allows us to identify specific LMXBs that are most likely to exhibit
signatures of the r-mode (in)stability.
We bring together the theoretical models that have been examined in the
past (involving, e.g., elasticity, exotic states of matter, and superfluidity)
and demonstrate that, based on our current state of knowledge,
they all fail to explain the observed systems.
This dilemma is irrespective of the Cassiopeia~A superfluid results.
We discuss the uncertainties and outline where advancements can be made.

{\it Neutron star core temperatures.}$-$
Let us assume that the r-mode instability is active in
LMXBs at the level required to balance the accretion torque,
while the associated heating is balanced by neutrino cooling.
The accretion luminosity $\Lacc$ and NS spin frequency $\nus$
($=\Omegas/2\pi$) are measured from LMXB observations.
Since the NS (with mass $M$ and radius $R$) is taken to be in
spin-equilibrium, the spin-up torque from accretion is equal to the
spin-down torque from gravitational radiation, i.e., $\Nacc=\Ngw$.
We take $\Nacc=\Lacc/\OmegaK$, where $\OmegaK=(GM/R^3)^{1/2}$ is the
Kepler rotation frequency, and
use the model of \cite{owenetal98,anderssonkokkotas01} to obtain $\Ngw$
for a r-mode with amplitude $\alpha$ and timescales $\tau$ for
relevant processes.
Considering a $1.4\,M_{\rm Sun}$, 12.5~km NS, the balance yields an
``equilibrium'' r-mode amplitude
\be
\alpha\approx 8\times 10^{-7}\!
 \left(\Lacc/10^{35}\mbox{ergs s$^{-1}$}\right)^{1/2}\!
 \left(\nus/300\mbox{ Hz}\right)^{-7/2}\!\!\!. \label{eq:alphaeq}
\ee
Our choice of $M$ and $R$ are in line with those used in previous r-mode
work \cite{owenetal98}, as well as those given by the
Akmal-Pandharipande-Ravenhall equation of state (see below).
In steady-state, the heat dissipated by damping of the r-mode
is equal to the energy gain from GW emission,
i.e., $\Lheat=-\Lgw$, where $\Lgw=-\Ngw\Omegas/3$,
so that \cite{brownushomirsky00} 
\be
\Lheat = \Lacc\Omegas/3\OmegaK = 0.065(\nus/300\mbox{ Hz})\Lacc.
 \label{eq:lheat}
\ee

\begin{figure}
\resizebox{0.95\hsize}{!}{\includegraphics{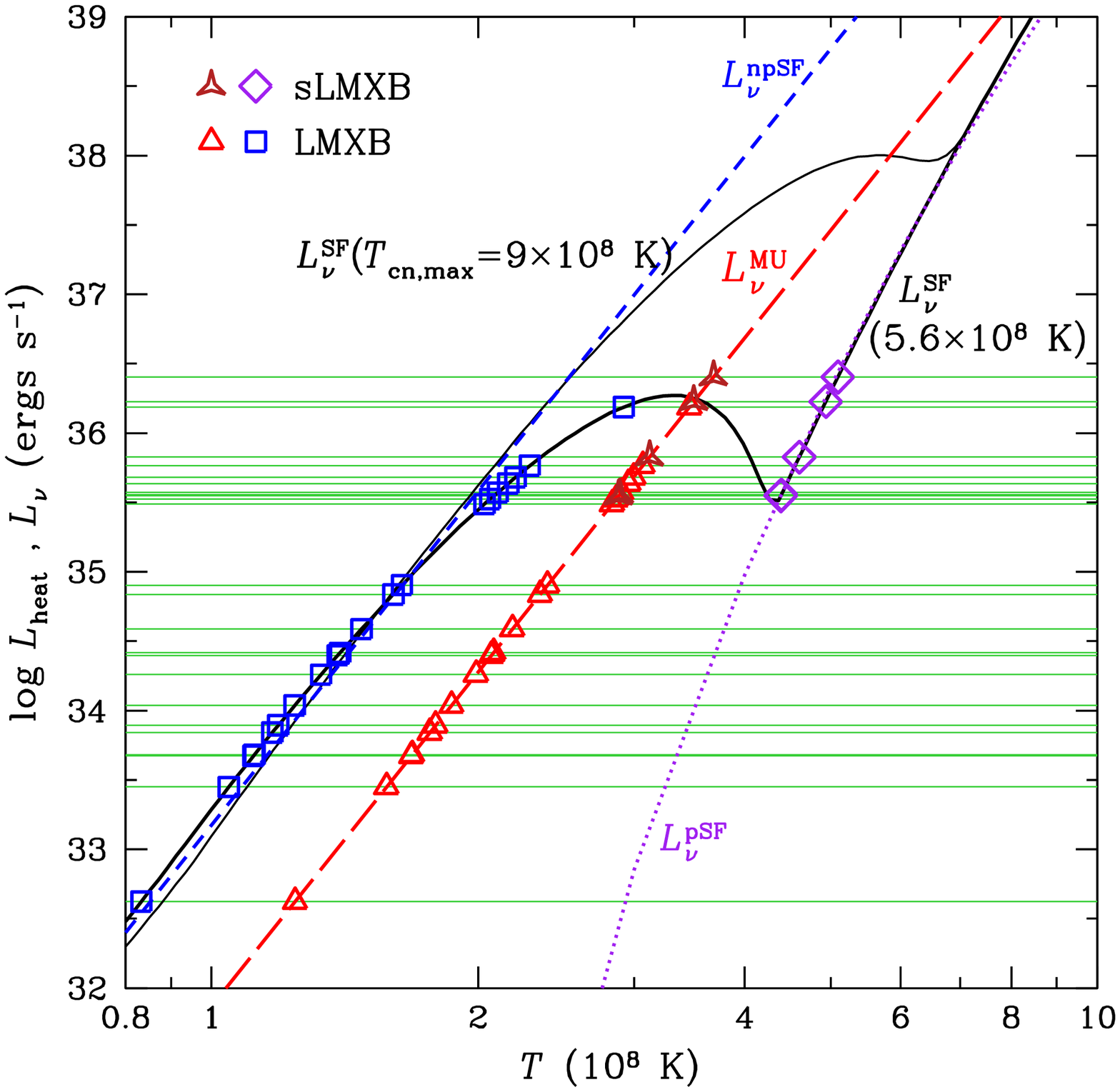}}
\caption{
Heat generated by damping of r-modes $\Lheat$ compared to the neutrino
cooling luminosity $\Lnu$ as a function of NS core temperature $\Tc$.
The thin horizontal lines are $\Lheat$ for known LMXBs computed using their
flux, distance, and spin frequency from \cite{wattsetal08}
and Eq.~(\ref{eq:lheat}).
The long-dashed line is the modified Urca luminosity $\Lmu$.
The triangles and starred-triangles indicate the intersection of $\Lheat$ and
$\Lmu$, which determines $\Tc$ for each LMXB and
short recurrence time LMXB (sLMXB).
The thick solid lines are $\Lsf$ with $\Tcnmax=5.6$ and $9\times 10^8$~K,
and the squares and diamonds are the inferred $\Tc$ (from $\Lheat=\Lsf$)
for each source.
The short-dashed and dotted lines are approximate fits to $\Lsf$ in the
strongly superfluid and in the non-superfluid neutron regimes, respectively.
}
\label{fig:ltemp}
\end{figure}

Taking the heat from r-mode dissipation to be lost by neutrino emission
[$\Lheat=\Lnu(\Tc)$], the core temperature $\Tc$ can be inferred.
Note that cooling via neutrino emission dominates over photon emission
at our considered temperatures.
It is traditional to assume that the NS cools by the modified Urca
neutrino emission process, which has a luminosity \cite{shapiroteukolsky83}
\be
\Lmu\approx 7.4\times 10^{31}\mbox{ ergs s$^{-1}$ }(\Tc/10^8\mbox{ K})^8.
 \label{eq:lmu}
\ee
Setting $\Lheat$ equal to $\Lmu$ yields the core temperature.
Figure~\ref{fig:ltemp} shows $\Lheat$ and $\Lmu$,
with their intersection indicating the core temperature for each LMXB.
It is worth noting that the heat associated with the unstable r-mode
[Eq.~(\ref{eq:lheat})] corresponds to $\sim 10$~MeV per accreted nucleon,
compared to the $\sim 1$~MeV from nuclear burning in the deep crust
\cite{haenselzdunik90}.
Even if we assume nuclear heating instead of an unstable r-mode,
the effect on the inferred $\Tc$ is only at the $\lesssim$~30\% level
because of the strong temperature scaling in Eq.~(\ref{eq:lmu}).

%-------------------------------------------
%-------------------------------------------

The above estimates assume normal nucleons in the stellar interior.
It is expected that neutrons are superfluid and
protons are superconducting in the NS core \cite{haenseletal07}.
The measurement of rapid cooling of the Cassiopeia~A NS
\cite{heinkeho10,shterninetal11} gives the first
direct evidence for the existence of superfluid components and constrains
the critical temperatures for the superfluid transition $\Tcn$ and $\Tcp$,
i.e., $\Tcnmax\approx(5-9)\times 10^8$~K and
$\Tcp\sim (2-3)\times 10^9$~K \cite{pageetal11,shterninetal11}.
Superfluidity has two important effects on neutrino emission and cooling:
(1) suppression of emission mechanisms, like the modified Urca process,
that involve superfluid constituents and
(2) enhanced emission near the critical temperatures due to Cooper pair
formation \cite{yakovlevpethick04}.
We use the results of \cite{yakovlevetal99} to
calculate the neutrino emissivities due to the modified Urca process,
accounting for superfluid suppression, and the Cooper pair formation
process.
We take $\Tcp=2\times 10^9$~K and $\Tcn(\rho)$ to be approximately
given by model (a) of \cite{shterninetal11}.
The neutrino luminosity $\Lsf$ is then obtained by integrating the
emissivities using a stellar model based on the APR EOS with
$M=1.4\,M_{\rm Sun}$ and $R=12$~km \cite{shterninetal11}.
The results presented here do not depend strongly on the assumed stellar mass
\cite{pageetal11,shterninetal11}.

Figure~\ref{fig:ltemp} shows $\Lsf$ with $\Tcnmax=5.6$ and $9\times 10^8$~K.
At $\Tc\gtrsim\Tcnmax$, the suppression of the modified Urca process
by the superconducting protons yields $\Lsf<\Lmu$.
Since cooling is less efficient, the inferred core temperatures are
higher than those obtained from Eq.~(\ref{eq:lmu}).
At $\Tc<\Tcnmax$, neutrino emission is enhanced due to
Cooper pair formation, and the cooling is more efficient, which
results in a lower inferred $\Tc$.
It is noteworthy that, for $\Tcnmax\lesssim 8\times 10^8\mbox{ K}$,
a unique $\Tc$ does not exist for a range of $\Lsf=\Lheat$.
For example, for $\Tcnmax=5.6\times 10^8$~K, there can be a factor of two
difference in the inferred $\Tc$ when the observed accretion luminosity
$\Lacc\sim (3-10)\times 10^{36}\mbox{ ergs s$^{-1}$}(600\mbox{ Hz}/\nus)$.
Interestingly, there are five LMXBs that show short recurrence times
between multiple X-ray bursts due to nuclear burning of accreted matter
\cite{keeketal10}. These sources have accretion luminosities within this range,
and thus their higher temperatures could perhaps be responsible for their
distinct bursting behavior.
We also note that there is a branch of $\Lsf$ that could produce LMXBs
which increase in luminosity even though their temperatures are decreasing.

\begin{figure}
\resizebox{0.8\hsize}{!}{\includegraphics{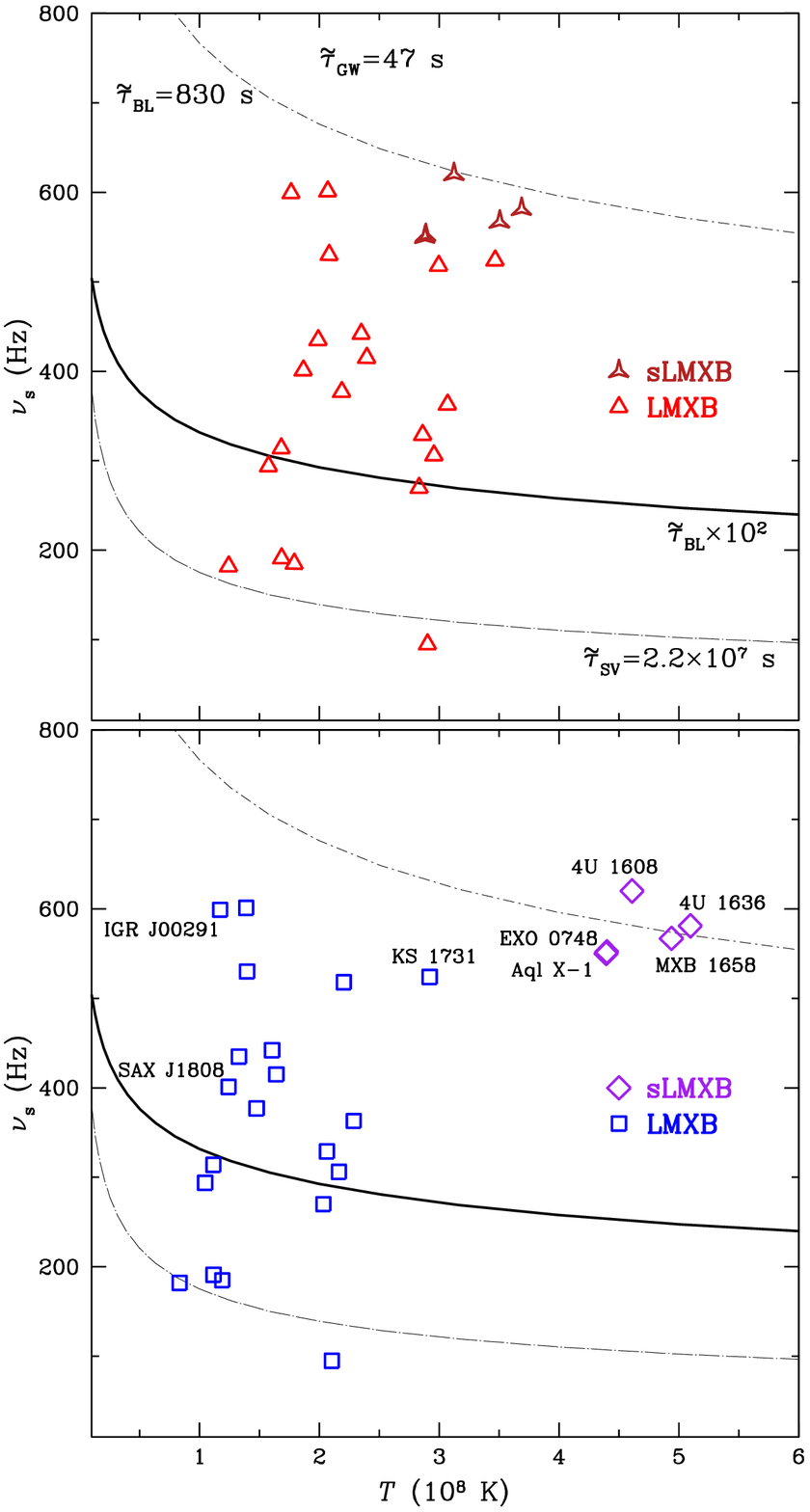}}
\caption{
Neutron star spin frequency $\nus$ and core temperature $\Tc$.
The long-dashed-dotted lines are the shear viscosity instability curve
(where $\tgw=\tsv$),
the short-dashed-dotted lines are the rigid boundary layer instability
curve (where $\tgw=\tek$),
and the solid lines are a more realistic, elastic boundary layer curve
(where $\tgw=10^2\tek$).
Top panel:
The triangles are LMXBs whose $\Tc$ are derived from their observed $\Lacc$
and assuming $\Lmu$ for cooling.
Bottom panel:
The squares are LMXBs and diamonds are short recurrence time LMXBs (sLMXB)
whose $\Tc$ are derived from their observed $\Lacc$
and assuming $\Lsf$ for cooling.
}
\label{fig:window}
\end{figure}

Finally, we find that, in the temperature regime ($\Tc\ll\Tcnmax$)
where both protons and neutrons are strongly superfluid, the neutrino
luminosity is \cite{yakovlevpethick04}
\be
\Lsflo \approx 20\Lmu, \label{eq:lsflo}
\ee
while in the temperature regime ($\Tcnmax\lesssim\Tc\ll\Tcp$) where protons
are superfluid and neutrons are normal, the neutrino luminosity is
\be
\Lsfhi\approx 4\times 10^{39}\mbox{ ergs s$^{-1}$}[\log(T/10^8\mbox{ K})]^{21}.
\label{eq:lsfhi}
\ee
Figure~\ref{fig:ltemp} also shows $\Lsflo$ and $\Lsfhi$.
By setting $\Lheat$ equal to $\Lsflo$ or $\Lsfhi$, we obtain core
temperatures which approximate the ones illustrated in Fig.~\ref{fig:window}.

%-------------------------------------------
%-------------------------------------------

{\it Physics of the instability window.}$-$
Figure~\ref{fig:window} shows the core temperature (inferred from either
$\Lheat=\Lmu$ or $\Lheat=\Lsf$) and spin frequency for each LMXB.
Since superfluidity suppresses damping mechanisms like hyperon bulk
viscosity and alternative mechanisms like mutual friction are too weak
(see below), the consensus view is that the viscous boundary layer at the
crust-core interface is the primary damping agent.
It is clear that a large number of LMXBs are in the
unstable region (above the $\tsv$-curve)
unless the damping is described by a rigid crust model
($\tek$-curve) \cite{bildstenushomirsky00}.
However, a rigid crust is completely at odds with expectations.
In the fast systems, the Coriolis force that drives the r-modes should
dominate the elastic restoring force
($\mu/\Omegas\sim 10^{-4}$, where $\mu$ is the shear modulus).
``Slippage'' between the crust and core reduces
the damping by a factor $>$100 (see Fig.~\ref{fig:window})
\cite{levinushomirsky01}.
It is also worth noting that the magnetic fields in these systems
($\sim 10^8$~G) are too weak to alter the nature of the boundary layer
(this requires core fields $\gtrsim 10^{11}$~G \cite{mendell01}).
We consider the implications of the data in Fig.~\ref{fig:window} in light
of these arguments.

First, let us assume that the r-modes are unstable.
One might expect the unstable systems to exhibit a distinctive behavior.
An example may be the short recurrence time LMXBs,
which would make them interesting targets for gravitational wave searches;
we estimate that dissipation from an unstable r-mode can
power the observed quiescent luminosity of these higher temperature LMXBs
(c.f.\ \cite{brownushomirsky00}).
Conversely, the low temperature LMXBs may be r-mode stable;
this idea is supported by the LMXBs SAX~J1808.4$-$3658 and IGR~J00291$+$5934,
which have measured spin evolutions that are consistent with magnetic dipole
losses without gravitational radiation \cite{falangaetal05};
note that the low temperature LMXBs could have even lower temperatures, if,
e.g., fast neutrino cooling processes operate in these sources
\cite{yakovlevpethick04}.
Consider a NS that enters the unstable region.
The r-mode then grows rapidly to an amplitude such that nonlinear
coupling to other modes causes the instability to saturate \cite{arrasetal03};
the saturation amplitude is expected to be much larger than 
that required for spin-balance [c.f.\ Eq.~(\ref{eq:alphaeq})].
The subsequent evolution is likely to be quite complex \cite{bondarescuetal07}.
In principle, the NS will heat up and spin-down, and the LMXB should leave
the instability window in a time much shorter than the age of the system
\cite{levin99}.
Therefore the observed LMXBs should {\it all} be stable,
which contradicts the data in Fig.~\ref{fig:window}.
Most importantly, all reasonable evolutionary scenarios
\cite{levin99,bondarescuetal07} predict maximum NS spin rates that are
far below those observed.

\begin{figure}
\resizebox{\hsize}{!}{\includegraphics{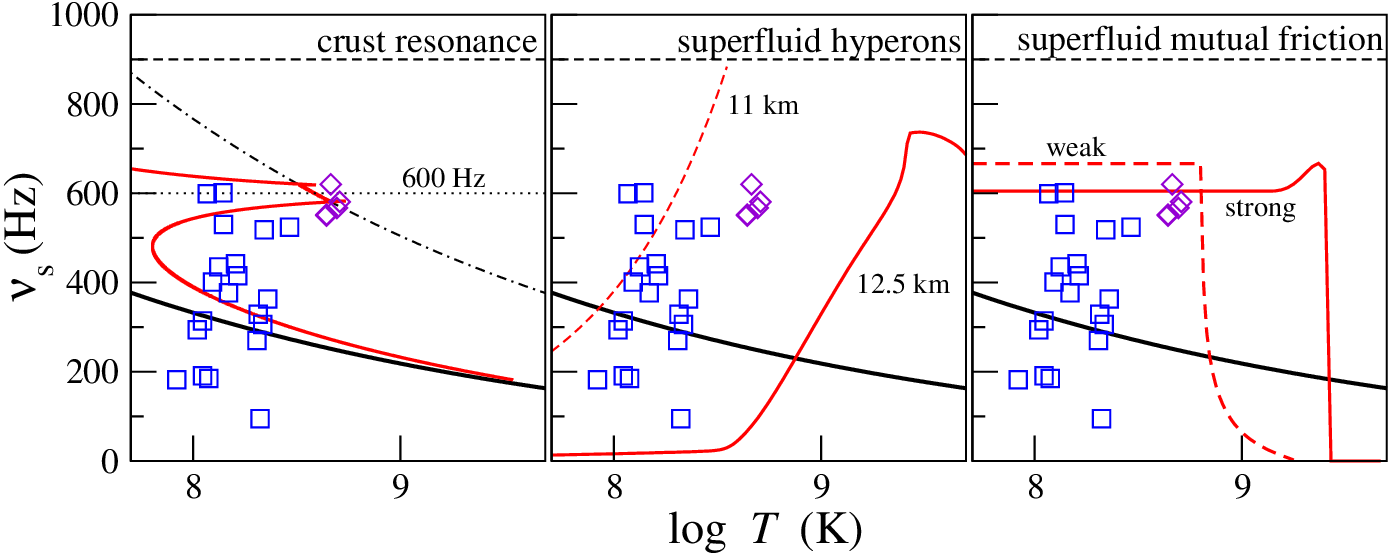}}
\caption{
Three scenarios that could explain r-mode stability in the observed LMXBs.
Left panel: Crust mode resonance at 600~Hz.
Middle panel: Superfluid hyperons (based on \cite{haskellandersson10}
with $\chi=0.1$).
Right panel: Strong vortex mutual friction
(based on the strong/weak superfluidity models from \cite{haskelletal09}
with $\mathcal{B}\approx 0.01$).
The dashed lines indicate the break-up limit.
}
\label{fig:windowphys}
\end{figure}

For r-mode stability, a revision of our understanding
of the relevant damping mechanisms is required.
We consider possible resolutions, starting with the viscous boundary layer.
The crust-core transition may be more complex than has been
assumed thus far.
This should be expected given the  presence of a type-II
superconductor in the outer core of the star \cite{haenseletal07}.
The details of the transition are likely to strongly affect the instability
window, but the problem has not attracted real attention.
Crust physics may also be vital.
There may be resonances between the r-mode and torsional oscillations of
the elastic crust \cite{levinushomirsky01}.
Such resonances would have a sizeable effect on the slippage factor,
leading to a complicated instability window.
Figure~\ref{fig:windowphys} gives an example;
the illustrated instability window has a relatively broad resonance
at 600~Hz, which is the typical frequency of the first overtone of
pure crustal modes.
Although our example is phenomenological (c.f.\ \cite{levinushomirsky01}),
it suggests that this mechanism may explain the stability of LMXBs.
Realistic crust models are needed to establish to what extent
this is viable.

%-------------------------------------------
%-------------------------------------------

Another possibility is an instability window that increases with temperature
\cite{anderssonetal02}.
If this is the case, then LMXBs may evolve to a quasi-equilibrium
where the r-mode instability is balanced (on average) by accretion and
r-mode heating is balanced by cooling (as in our temperature estimates).
This solution is interesting because it predicts persistent 
(low-level) gravitational radiation.
Figure~\ref{fig:windowphys} shows a model using hyperon bulk viscosity
suppressed by superfluidity.
However, this explanation has a major problem.
We must be able to explain how the observed millisecond radio pulsars
emerge from the accreting systems.
Once the accretion phase ends, the NS will cool, enter the instability
window, and spin down to $\sim 300$~Hz (see Fig.~\ref{fig:windowphys}).
In other words, it would be very difficult to explain the formation of
a 716~Hz pulsar \cite{hesselsetal06}.

A more promising possibility involves mutual friction due to vortices
in a rotating superfluid. The standard mechanism (electrons scattered off
of magnetized vortices) is too weak to affect the
instability window \cite{lindblommendell00}.
However, if we increase (arbitrarily) the strength of this mechanism by
a factor $\sim 25$,
then mutual friction dominates the damping (see Fig.~\ref{fig:windowphys}).
Moreover, this would set a  spin-threshold for instability similar to the
highest observed $\nus$ and would allow systems to remain rapidly rotating
after accretion shuts off.  Enhanced friction may result from the
interaction between vortices and proton fluxtubes in the outer core,
as proposed in a model for pulsar free precession \cite{link03}.
This mechanism has not been considered in the context of neutron star
oscillations and instabilities, but it seems clear that such work is needed.

In summary,
we considered astrophysical constraints on the r-mode instability
provided by the observed LMXBs.
Having refined our understanding of the likely core temperatures in these
systems using recent superfluid data, we showed that several systems lie
well inside the expected instability region.
This highlights our lack of understanding of the physics of the
instability and the associated evolution scenarios and at the same time 
points to several interesting directions for future work.

\begin{acknowledgments}
WCGH thanks Peter Shternin for the APR EOS.
WCGH appreciates use of computer facilities at KIPAC.
WCGH and NA acknowledge support from
STFC in the UK.
BH holds a EU Marie Curie Fellowship.
\end{acknowledgments}
%%%
\vspace{-0.4cm}
%%%%%%%%%%%%%%%%%%%%%%%%%%%%%%%%%%%%%%%%%%%%%%%%%%%

\end{document}